\documentstyle[12pt,a4]{article}

\newcommand{\beq}{\begin{equation}}
\newcommand{\eeq}{\end{equation}}
\newcommand{\bea}{\begin{eqnarray}}
\newcommand{\eea}{\end{eqnarray}}

\newcommand{\Ds}{\not\!\! D}

\newcommand{\aas}{/\kern-.50em A}

\newcommand{\ra}{\rightarrow}

\newcommand{\D}{{\cal{D}}}

\begin{document}


\title
{Chiral condensates and QCD vacuum in two dimensions}
\author{{\normalsize H. R. Christiansen} 
\thanks{Electronic address: hugo@cat.cbpf.br}\\
{\normalsize\it Centro Brasileiro de Pesquisas Fisicas. 
CBPF - DCP}\\
{\normalsize\it Rua Xavier Sigaud 150, 22290-180 
Rio de Janeiro, Brazil.}}

\maketitle

\begin{abstract}
\normalsize\noindent
We analyze the chiral symmetries of flavored quantum chromodynamics
in two dimensions and show the existence of chiral condensates
within the path-integral approach. The massless and massive cases
are discussed as well, for arbitrary finite and infinite number
of colors. 
Our results put forward the question of topological issues 
when matter is in the fundamental representation of the gauge group.
\end{abstract}


\section{Introduction \label{sec-intro}}

Correlation functions are key quantities in the understanding
of nonperturbative QCD and hadron physics. Basically, because the same 
functions can be considered in terms of fundamental QCD fields
or in terms of physical intermediate states. Using hadronic phenomenological
data, one can extract valuable information of the underlying 
structure of the theory. For example, the difference between vector
and axial correlators is enterely due to the chiral assymetry of the 
QCD vacuum \cite{shuryak}.
Thus, in order to get some insight into the complex structure of the 
QCD gound state one should consider possible mechanisms for chiral 
symmetry breaking and condensate formation. 
In general, these correlators are defined as the vacuum 
expectation values (v.e.v.) of the product of operators in different 
space-time points.
A nonvanishing v.e.v. of quarks composites 
can be understood  as a result of the condensation of pairs of particles 
and holes.   
The strong attractive interaction of quarks
and the low energy cost of creating a massless pair are responsible
for the appearence of a condensate. 
Further, the Fourier transform
of correlators shows that there is a momentum transfer flowing between
the composite operators, making apparent the connection between them
by means of the vacuum fields.
On the other hand, conservation of the quantum 
numbers of the vacuum implies a net chiral charge of such correlators. 
Hence, these quantities
are particularly useful in connection to
both spontaneous and dynamical chiral symmetry breaking, 
as well as to shed some light on crucial aspects of quantum chromodynamics
such as its topological structure.

Two-dimensional models like quantum chromodynamics in 
two dimensions (QCD$_2$),
are convenient frameworks to discuss these kind of phenomena since
they present the basic aspects of the fundamental four dimensional 
theories (such as 
the existence of nontrivial topological sectors and chirality 
properties) and, furthermore, analytical results can be 
generally obtained. 

In the present work we analyze the chiral symmetries of the 
QCD$_2$ ground state by means of fermionic local correlators.
Using a path-integral approach which is very appropriate
to handle non-Abelian gauge theories,
we calculate  vacuum expectation values
of products of local bilinears $\bar\psi(x) \psi(x)$ 
in two-dimensional quantum chromodynamics with flavor.  
In this framework, we show that certain
multipoint chiral condensates are nonvanishing, a result which
signals the dynamical breakdown of the full $U(N_f)\times U(N_f)$ chiral 
symmetry group down to a smaller subgroup. 
However, the elementary mass term has a zero v.e.v. in the massless model, 
a result which is compatible 
with a vanishing isosinglet chiral anomaly and which is consistent with 
Coleman's theorem \cite{colth}. Thus, the isosinglet axial current is
conserved in two-dimensional massless QCD. 

In this respect, we address the existence of an elementary
chiral condensate in the large $N_c$ limit, reported in 
Refs. \cite{zhit,lenz,mi}.
To connect our path-integral approach to these alternative 
procedures, we work over the massive theory and
perform a series expansion in the fermionic mass
so as to argue about the topological origin of this nontrivial mass like
condensate appearing in the 't Hooft's weak phase.
The topological structure of the theory is especially considered
and we show the central role played by topologically charged sectors in 
obtaining nonzero correlators.

\section{Topology}

In two space-time dimensions it is generally assumed that the 
vanishing of the 
homotopy group $\pi_1 (SU(N))$ implies that QCD$_2$ exhibits
only  a trivial topology  when fermions are in the fundamental
representation of the gauge group. Hence, no vacuum degeneracy 
is expected to occur \cite{smilga}.
In contrast, when adjoint matter is considered, the relevant 
symmetry group becomes $SU(N)/Z_N$ rather than $SU(N)$. This
implies that $\pi_1\neq 0$ which
is responsible for the appearence of $N$ topologically different 
sectors and instanton effects become apparent.
 
Nevertheless, by handling fundamental fermions it can be 
easily verified 
that gauge field configurations lying in the Cartan subalgebra of 
$SU(N)$ generate non\-trivial topological 
fluxes. Thus, one should
include these topologically charged configurations in the path-integral
{\em also} for fermions in the fundamental 
represen\-tation. 
It should be noticed however,
that while $Z_N$ is the relevant homotopy group for fermions in both 
the fundamental or the adjoint 
representation of the gauge group, instantons are not solutions of QCD$_2$
in any case.
Indeed, in two Euclidean dimensions one needs scalar fields and 
spontaneous symmetry breaking in order to have finite action solutions.
Precisely, one has to arrange scalars so as to break the symmetry
of the original gauge group down to its center $Z_N$ since only in this case
topologically nontrivial solutions (classified in homotopy classes
associated with the elements of $\pi_1 (SU(N)/Z_N)=Z_N$) 
do exist, independently of the 
fermion representation
one will afterwards choose when adding matter.

Concerning this point, let us stress that
in two dimensions the role of instantons
is basically played by vortices. In the Abelian case, these vortices
are identified with the Nielsen-Olesen vortex solutions
of a spontaneously broken Abelian Higgs model \cite{NO}. 
This should be contrasted with real QCD where four-dimensional 
instantons are regular solutions of the gauge field
equations of motion when fermions are absent; as we stated above,
in the two-dimensional case, either Abelian or
non-Abelian, no regular solutions with topological charge exist 
unless complete symmetry breaking is achieved via Higgs fields. 
When these scalars are included, the resulting static, axially
symmetric gauge field configurations  give a
realization in a two-dimensional Euclidean theory, 
of regular gauge fields carrying a topological charge.
These classical configurations  
are then identified with two-dimensional instantons,
to be used in non-perturbative analysis of a Maxwell theory
coupled to massless fermions \cite{Col}. 
The same route can be undertaken
in the non-Abelian case since the analogous to Nielsen-Olesen
vortex solutions have been shown to exist, again for
spontaneously broken gauge theories \cite{dVS}.
It means that, in the spirit of the path-integral approach to 
quantum field theory
one cannot forget to include these configurations in the integration 
domain of the theory-without Higgs fields.  
Once regular gauge field configurations carrying
topological charge are identified, the associated fermion
zero modes can be found \cite{JR,dV} and then, 
used to study the formation of fermion condensates. 

In order to achieve the path-integration  it would be worth performing
a decoupling transformation of gauge fields from fermions.
It is important that the decoupling operation
does not change the fiber 
bundle in which the Dirac operator is defined, hence
we start by decomposing every gauge field belonging 
to the $n^{th}$ topological sector in the form \cite{bc}
\beq
A_\mu^a(x) = A_\mu^a{}^{(n)} + a_\mu^a
\label{pi}
\eeq
where $A_\mu^{(n)}$ is a 
classical fixed configuration of $n^{th}$ 
flux class and $a_\mu$ is the path-integral variable which takes into
account quantum fluctuations. The quantum field $a_\mu$ belongs
to the trivial topological sector and can then be decoupled from fermions
by a chiral rotation yielding a Fujikawa jacobian \cite{fuji}. Thus, the
integration measure must be only defined on the $n=0$ sector.

Topological gauge field configurations and
the corresponding zero-modes of the Dirac equation play a central
role in calculations involving fermion composites. 
As in the Abelian case, two-dimensional gauge field configurations
$A_\mu^{(n)}$ carrying a topological charge $n \in Z_N$ can be found 
for the $SU(N)$ case. The relevant homotopy group in this case is
$Z_N$ and not $Z$ as in the $U(1)$ case \cite{dVS2}.
Taking $g_n$ in the
Cartan subgroup of the gauge group
we can write a gauge field configuration belonging to the $n^{th}$ 
topological sector in the form
\beq
A_\mu^{(n)} = i A(\vert z\vert )\ g_n^{-1} \partial_\mu g_n
\label{gf}
\eeq
with $A(0)=0$ and $\lim_{|z|\ra\infty} A(|z|)=-1$, where $z = x_0 + i x_1$.

Zero modes of the Dirac operator in the background of
such non-Abelian vortices,  have been analyzed 
in \cite{dV}.
The outcome is that for topological charge $n>0$ ($n<0$) there are $Nn$ 
($N\vert n \vert$)
square-integrable zero modes $\eta_L$ ($\eta_R$) analogous to those
arising in the Abelian case. Indeed, one has
\beq
\eta_R^{(m,i) j} = \left(\begin{array}{c} z^m h_{ij}(z,\bar z) \\ 0
\end{array} \right),
\ \ \ \
\eta_L^{(m,i) j} = 
\left(\begin{array}{c} 0 \\{\bar z}^{-m} h_{ij}^{-1}(z,\bar z)
\end{array} \right)
\label{naz}
\eeq
with
\beq
h(z,\bar z) = \exp(\phi^{(n)}(\vert z\vert) M),
\ \ \ \ 
M = \frac{1}{N} {\rm diag} (1,1, \ldots, 1-N)
\label{cui}
\eeq
and $\phi^{(n)}$ is given by
\beq
\epsilon_{\mu \nu} \frac{x_\nu}{\vert z \vert}
\frac{d}{d\vert z\vert }\phi^{(n)}(\vert z \vert) =A_\mu^{(n)}. 
\label{ff}
\eeq
Here $i,j= 1,2, \ldots, N$ and $m = 0,1, \ldots, \vert n \vert - 1$. 
The pair $(m,i)$ labels the $N\vert n \vert$ different zero-modes while
$j$ corresponds to a color projection index. 


\section{Fundamental matter in $QCD_2$}

Let us consider two dimensional $SU(N_c)$ Yang-Mills gauge fields 
coupled to massless Dirac fermions in the fundamental representation
of the group in Euclidean space-time
\beq
L=\bar\psi^{q}(i\partial_{\mu} \gamma_{\mu} \delta^{qq'}+A_{\mu,a}
 t_a^{qq'}\gamma_{\mu})\psi^{q'}+
\frac{1}{4g^2} F_{\mu\nu}^a F_{\mu\nu}^a.
\label{lag}
\eeq
Here the labels $a=1\dots N_c^2-1,\ $ and $q=1\dots N_c$ are summed over,
and the partition function is
\beq
Z = \int \D \bar\psi \D \psi\D A_\mu \exp[-\int d^2x\,  L]. 
\label{Z}
\eeq

In order to compute fermionic correlators 
containing products of local bilinears $\bar \psi \psi(x)$ 
it will be convenient to decouple fermions from the $a_{\mu}$ field
through a chiral rotation within the topologically trivial sector. 
The choice of an appropriate background like
\beq
A^{a (n)}_+ = 0
\label{ba}
\eeq
is important in order to control the zero-mode problem \cite{multi}.
Let us start by introducing
group-valued fields to represent $A^{(n)}$ and $a_\mu$
\beq
a_+ = i u^{-1} \partial_+ u
\label{u}
\eeq
\beq
a_- = i d(v \partial_- v^{-1}) d^{-1}
\label{vv}
\eeq
\beq
A^{(n)}_- = i d \partial_- d^{-1}.
\label{a}
\eeq
In terms of these fields the fermion determinat can be suitably 
factorized in an arbitrary gauge
by repeated use of the Polyakov-Wiegmann identity \cite{polw},
resulting in
\beq
\det \Ds[A^{(n)} + a] = 
{\cal N} \det \Ds[A^{(n)}] e^{-S_{eff}[u,v; A^{(n)}]}
\label{sui}
\eeq
where
\begin{eqnarray}
S_{eff}[u,v; A^{(n)}] & = &
W[u, A^{(n)}]+ W[v] + \frac{1}{4\pi}tr_c\!\int\! d^2x\, 
(u^{-1} \partial_+ u) d (v \partial_- v^{-1}) d^{-1}
 \nonumber \\
& & +\frac{1}{4\pi}tr_c\!\int\! d^2x\, (d^{-1} \partial_+ d) 
(v \partial_- v^{-1}).
\label{pris}
\end{eqnarray}
Here $W[u,A^{(n)}]$ is the gauged Wess-Zumino-Witten action which
in this case takes the form
\beq
W[u, A^{(n)}] = W[u] + \frac{1}{4\pi}tr_c\int d^2x (u^{-1} 
\partial_+ u) (d \partial_- d^{-1})
\label{sa}
\eeq
and $W[u]$ is the usual WZW action.

Once the determinant has been written in the form (\ref{sui}),
one can work with any gauge choice. The partition function shows 
the following structure 
\bea
Z  &=&  \sum_n  \det(\Ds[A^{(n)}])  \int \D a_\mu\,
\Delta_{FP}\, \delta(F[a])\nonumber\\  
& & \exp \left( -S_{eff}[A^{(n)}, a_\mu] - \frac{1}{4g^2} 
\int d^2x  F^2_{\mu\nu}[A^{(n)}, a_\mu] \right)
\label{z1}
\eea
where $\Delta_{FP}\, \delta(F[a])$ comes from the gauge fixing.


\subsection*{Condensates of fundamental matter fields }

As it happens in the Abelian case, the partition function of two
dimensional quantum chromodynamics
only picks a contribution from the trivial sector because
$\det(\Ds[A^{(n)}])=0$ for $n\neq 0$ (see eq.(\ref{z1})). 
In contrast, various correlation functions become nontrivial precisely
for $n\neq 0$ thanks to the  zero-mode contributions when 
Grassman integration is performed.

In order to obtain a general expression for multipoint local correlators
let us  work with the gauge choice given in eq.(\ref{ba}). Now,
the Dirac equation takes the form
\beq
\Ds[A^{(n)} + a]\left( \begin{array}{c}
			\psi_+ \\ 
			\psi_-
			\end{array} \right) =
 \left( \begin{array}{cc} 0 & u^{-1}i\partial_+  \\
 dvd^{-1}D_-[A^{(n)}]	 & 0 \end{array} \right)
\left( \begin{array}{c}
			\zeta_+ \\ 
			\zeta_-
			\end{array} \right)
\label{matrix2}
\eeq
where $\zeta$ is defined by
\beq
\psi_+=dvd^{-1}\zeta_+,\ \ \ \psi_-=u^{-1}\zeta_-
\label{lasttrafo}
\eeq
Thus, the interaction Lagrangian in the $n^{th}$ flux sector 
reads
\beq
L =\bar\psi\Ds[A^{(n)}+a]\psi= 
\zeta_+^*\Ds_-[A^{(n)}]\zeta_+ + \zeta_-^*i\partial_+\zeta_-
\label{Lzeta}
\eeq
which we will write as 
$\ \bar\zeta \widetilde D[A^{(n)}]\zeta$.
In terms of these new fields, the elementary bilinear $\bar\psi\psi$
takes the form
\beq
\bar\psi\psi=\zeta_-^*u dvd^{-1}\zeta_+ +\zeta_+^* 
dv^{-1}d^{-1}u^{-1}\zeta_- .
\eeq
Notice that the fermionic jacobian associated with eq.(\ref{lasttrafo}) 
is just the effective action defined in the previous section 
by eq.(\ref{pris}).
Hence, arbitrary non-Abelian correlators of fundamental fermions
can be put down as
\bea
& & \langle \bar\psi\psi(x^1)\dots \bar\psi\psi(x^l)\rangle=
\sum_n\int \D u\D v\ \Delta_{FP}\ \delta (F[a_{\mu}])\, 
\exp[ -S_{eff}(A^{(n)},u,v) ]
\nonumber\\
& & 
\int \D\bar\zeta \D\zeta\ \exp( \bar\zeta 
\left( \begin{array}{cc} 0 & i\partial_+  \\
D_-[A^{(n)}] & 0 \end{array} \right)\zeta )\nonumber\\
& & 
\left( B^{q_1p_1}(x^1)\dots B^{q_lp_l}(x^l)\
\zeta_-^{*q_1}\zeta_+^{p_1}(x^1)\dots\zeta_-^{*q_l}\zeta_+^{p_l}(x^l)
+B^{q_1p_1}(x^1)\dots \right.\nonumber\\
& & 
B^{-1 q_lp_l}(x^l)\ \zeta_-^{*q_1}\zeta_+^{p_1}(x^1)\dots 
\zeta_+^{*q_l}\zeta_-^{p_l}(x^l)
+B^{q_1p_1}(x^1)\dots  \nonumber\\
& & 
B^{-1 q_{l-1}p_{l-1}}(x^{l-1})B^{-1 q_lp_l}(x^l)\
\zeta_-^{*q_1}\zeta_+^{p_1}(x^1)\dots 
\zeta_+^{*q_{l-1}}\zeta_-^{p_{l-1}}(x^{l-1})
\zeta_+^{*q_l}\zeta_-^{p_l}(x^l)
\nonumber\\
& & 
\left. +\dots + 
B^{-1 q_1p_1}(x^1)\dots B^{-1 q_lp_l}(x^l)\
\zeta_+^{*q_1}\zeta_-^{p_1}(x^1)\dots\zeta_+^{*q_l}\zeta_-^{p_l}(x^l)
\right)
\label{grande}
\eea
where the  group-valued field $B$ is given by $B=u dvd^{-1}$.
Note that this is a general and completely 
decoupled expression for fermionic correlators,  which shows 
that the simple product found in the Abelian case becomes here 
an involved sum due to color couplings.

The introduction of a flavor index implies additional degrees 
of freedom which result in $N_f$ independent fermionic field variables. 
Consequently, the growing number of Grassman (numeric) differentials 
calls for additional Fourier coeficients in the integrand. 
Dealing with $N_f$ fermions coupled to the gauge field,  
leads to the fermionic jacobian computed for one
flavor to the power $N_f$, the bosonic measure remaining the same.
As we have previously explained, the Dirac operator 
has $|n|N_c$ zero modes in the $n^{th}$ topological sector, 
implying that more fermion bilinears will be needed in order to obtain 
a nonzero fermionic path-integral.  Moreover, 
since flavor comes together with a factor $N_f$ on the number of
Grassman coeficients, the minimal nonzero
product of fermion bilinears 
in the $n^{th}$ sector requires of $|n|N_c N_f$ insertions. 

Since the properties of all these topological
configurations are given by those in the torus 
of $SU(N_c)$, one can easily extend the results 
obtained in the Abelian case. In particular,
the chirality of the zero modes is dictated by the same index
theorem found in the Abelian theory, this implying that in sector $n>0$
($n<0$) every zero mode has positive (negative) chirality. In this way,
the right (left) chiral projections of the minimal nonzero 
fermionic correlators can be easily computed.

Since equations became a bit involved, in order to illustrate simple
flavored extensions of expression (\ref{grande}),
we just consider $N_f=2$ for two colors. 
The minimal fermionic correlator then  looks  
\bea
& & \sum_n\langle\bar\psi^{1,1}_+\psi^{1,1}_+(x^1)
\bar\psi^{1,2}_+\psi^{1,2}_+(x^2)
\bar\psi^{2,1}_+\psi^{2,1}_+(y^1)
\bar\psi^{2,2}_+\psi^{2,2}_+(y^2)\rangle{}_n=\nonumber\\
& &  \frac{1}{Z^{(0)}}\sum_{p,q,r,s}^{N_c=2}
\prod_{k=1}^2 \! \int_{gf}\!\!\!\! \D u \D v J_B\; 
e^{-S_{Beff}^{(1)}(u,v,d)}\, B^{1, p q}_k(x^k) B^{2, r s}_k(y^k) 
\times \nonumber\\
& &   
\int \D\bar\zeta_k \D\zeta_k\, 
e^{\int \bar\zeta_k \widetilde D[A^{(1)}]\zeta_k}\
\bar\zeta_+^{p,k}\zeta_+^{q,k}(x^k)
\ \bar\zeta_+^{r,k}\zeta_+^{s,k}(y^k).
\label{nabex}
\eea  
where 
\beq
B{}_k^{q, p_i l_i}(x)= u^{p_i q}(x) (dvd^{-1})^{q l_i}(x),
\eeq
$\widetilde D[A^{(n)}]$ is
the Dirac operator as defined in eq.(\ref{Lzeta}),
and $\bar\zeta_+ $ stands for $\zeta_-^*$.
We have employed the notation $Z^{(0)}$ for the partition function to
emphasize that
it is completely determined within the $n=0$ sector, see 
eq.(\ref{z1}). 
 The $gf$ 
subindex stands for the gauge fixing.
The action $S^{(n)}_{Beff}(u,v,d)=N_f S_{WZW}(u,v,d)+S_{Maxwell}(u,v,d)$ 
is given by the  
full gluon field $A^{(n)}(d)+a(u,v)$, and yields a high order 
Skyrme-type lagrangian \cite{fns}.

The fermionic path-integral can be easily performed, amounting
to a product of the eigenfunctions discussed in the sections above, 
as follows
\bea
& & \int \D\bar\zeta_k \D\zeta_k\ 
e^{\int \bar\zeta_k \widetilde D[A^{(1)}]\zeta_k}\ 
\bar\zeta_+^{p,k}\zeta_+^{q,k}(x^k)
\ \bar\zeta_+^{r,k}\zeta_+^{s,k}(y^k)= 
\det\prime(\widetilde D[A^{(1)}])\times\nonumber\\
& & \left( -\bar\eta_+^{(0,1)p,k}\eta_+^{(0,1)q,k}(x^k)
 \bar\eta_+^{(0,2)r,k}\eta_+^{(0,2)s,k}(y^k)
+\bar\eta_+^{(0,1)p,k}\eta_+^{(0,2)q,k}(x^k)\right.
\nonumber\\
& & 
\bar\eta_+^{(0,2)r,k}\eta_+^{(0,1)s,k}(y^k)
-\bar\eta_+^{(0,2)p,k}\eta_+^{(0,1)q,k}(x^k)
 \bar\eta_+^{(0,1)r,k}\eta_+^{(0,2)s,k}(y^k)\nonumber\\
& & \left.
+\bar\eta_+^{(0,2)p,k}\eta_+^{(0,2)q,k}(x^k)
 \bar\eta_+^{(0,1)r,k}\eta_+^{(0,1)s,k}(y^k)\right).
\label{ultima}
\eea 
Here $\det\prime(\widetilde D[A^{(1)}])$ is the determinat of the Dirac 
operator defined in eq.(\ref{Lzeta}) omitting zero modes, and (e.g.)
$\eta^{(0,1)q,k}(x^k)$ is a non-Abelian zero-mode as defined in section 2,
with an additional flavor index $k$.
Concerning the bosonic sector, the presence of the $F_{\mu\nu}^2$
(Maxwell) term crucially changes the effective dynamics with respect
to that of a pure Wess-Zumino model. One then has to perform 
approximate calculations  to compute the bosonic factor,
for example, by linearizing the $U$ transformation, see \cite{fns};
nevertheless, the point relevant to our discussion of
obtaining nonzero fermionic correlators is manifest in eq.(\ref{ultima}). 


\section{Nonzero $\langle\bar\psi\psi\rangle$}

As a byproduct, our approach gives $\langle \bar\psi\psi\rangle = 0$
in every flux sector including $n=0$. 
This is consistent with  Coleman's theorem
which prohibits the spontaneous breakdown of any continuos
symmetry in two dimensions. Furthermore, since, in contrast to the 
Abelian case, QCD$_2$ presents no (isosinglet) axial anomaly 
which could give 
rise to a nonzero mass like condensate $\langle\bar\psi\psi\rangle$. 

By the other side the existence of nonvanishing elementary condensates 
have been discussed within alternative scenarios for one flavor QCD$_2$
\cite{zhit,lenz,smilga}.
One is based on the assumption of an infinite number
of colors while the other focuses matter in the adjoint representation
(see also next section). This reminds one the outcome for massless 
two-dimensional QED \cite{jaye,hf}.

It is well known that in the Abelian case a multipoint composite
receives contributions from different topological sectors. In particular,
one can obtain the value of the elementary scalar condensate from 
a two point correlator
since the term $\langle \bar\psi_+\psi_+(x)\bar\psi_-\psi_-(y) \rangle$
factorizes as  $\langle \bar\psi_+\psi_+(x)\rangle \cdot
\langle\bar\psi_-\psi_-(y) \rangle$, the two factors being equal each other
and nonzero \cite{jaye}. Then, by means of the chiral decomposition
$\langle \bar\psi\psi(x)\rangle =\langle \bar\psi_+\psi_+(x)
\rangle+ \langle\bar\psi_-\psi_-(x) \rangle$ one constructs the  
condensate from the trivial topological sector although 
$\langle\bar\psi_{\pm}\psi_{\pm}(x)\rangle$ come exclusively from 
flux classes $\pm 1$  respectively \cite{hf}. The nonzero result 
indicates the breakdown of the chiral symmetry and it is known to 
be due to the $U(1)$ anomaly.

In the non-Abelian theory in turn, the result 
$\langle\bar\psi\psi\rangle=0$ thereby entails no anomalous chiral
symmetry in the singlet channel.
For a finite number of colors this is in agreement with 
independent analytical calculations based on operator product 
expansion and dispersion relations
\cite{zhit} and canonical quantization on the light-cone front
\cite{lenz}. 
Though, in contrast to ours, these  
previous analysis have been only 
performed in the trivial topological sector and, further, it has been  
assummed that cluster decomposition holds.

In the large $N_c$ limit, cluster decomposition takes place so that
all v.e.v. can be reduced to a product of elementary scalar densities,
i.e. $\langle A B \rangle = \langle A \rangle\langle B \rangle + O(1/N_c)$.
On the other hand, for very separated composites factorization also holds.
Outside any of these asymptotic situations we have shown which are the 
actual results for arbitrary values of both color and relative positions.
Notice that only for an infinite number of colors
the Berezinskii-Kosterlitz-Thoules (BKT)
behaviour \cite{bkt} of such an elementary 
fermion correlator  is compatible with a nonzero outcome 
for $\langle\bar\psi\psi\rangle$.  
The BKT effect can be shown 
by means of QCD sum rules giving a nonzero condensate
when the weak coupling regime is considered. In fact, bosonization 
rules show that for large  distances \cite{zhit}
\beq
\langle \bar\psi_+\psi_+(x)\bar\psi_-\psi_-(y)\rangle\sim |x-y|^{-1/N_c}
\label{bkt}
\eeq
implying that for large but finite $N_c$ it smears away softly.
Actually, this effect cannot be 
seen unless the $M \ra 0$ limit is taken at the end of the calculation
in the {\em massive} theory.
Moreover, the BKT behavior is a very especial result
that only takes place under 
the $g^2N_c=const$ condition, in a particular asymptotic direction. 
The limit  $N_c\ra \infty$, $g\ra 0$ and 
$M\ra 0$ is understood -but only- provided $M>>g\sim 1/\sqrt{N_c}\ra 0$
('t Hooft regime). 
Obviously this
establishes a very particular phase transition wherever one of the 
constraints does not
hold together with the others. Furthermore, the value of the 
correlator also depends on the choice of the approximation 
to the limit $N_c \ra \infty, \ |x-y|\ra \infty$, see eq.(\ref{bkt}).


\section{The massive case}

As we have mentioned, this BKT phenomenom only takes place provided the 
chiral symmetry is explicitly broken from 
the beginning by means of a quark mass term. 
One could then say that after the chiral weak phase is reached
in the massive theory,
the symmetry then keeps in a broken phase, but instead, 
it happens dynamically.
The physics changes,  the vacuum  being dressed non-perturbatively 
by means of purely planar diagrams \cite{largeN}. 
The spectrum is completely different depending on the relations 
shown above;
in the the weak phase there is an infinite number of massive mesons
while in the strong coupling phase there are just massless baryons.
In the massless non-Abelian theory the `pion' is 
also massless but there is no spontaneous symmetry breaking as dictated 
by Coleman's theorem.
Hence, in order to connect this nonzero outcome for the elementary 
condensate, we should extend our procedure to massive QCD$_2$.

Now, the partition function can be put as
\beq
Z = \int \D \bar\psi \D \psi\D A_\mu \exp(-\int d^2x\,  L_{M=0})
\ e^{-M \int d^2x\, \bar\psi\psi}. 
\label{zm}
\eeq
Actually, the analytical solution of this model is still lacking
and it might be a fruitless effort to attack this problem.
Nevertheless, for our purposes only small masses need to be examined.
Written in this form, it is suggested that we may perform a perturbative 
expansion in terms of the quark mass \cite{mass}. In 
order to compare the outcome with the 
results above let us  consider one flavor fermions. 
Then, we can proceed with the
minimal `massive' condensate as follows
\beq
\langle\bar\psi\psi(\omega)\rangle_M =\sum_n \int \D \bar\psi 
\D\psi\D a_\mu 
\exp(-\int d^2x\,  L^{(n)}_{(M=0)})\ \bar\psi\psi(\omega) e^{-M 
\int d^2x\, \bar\psi\psi}. 
\label{cm}
\eeq
In this fashion, it is apparent that for $M\neq 0$ this elementary 
condensate
receives contributions from every correlator coming from the massless
theory
\bea
& & \langle\bar\psi\psi(\omega)\rangle_M =\sum_n
\langle\bar\psi\psi(\omega)\rangle_{M=0}^{(n)} + 
M\ \sum_n \int d^2x\, \langle\bar\psi\psi(\omega)\bar\psi\psi(x)
\rangle_{M=0}^{(n)} +
\nonumber\\
& & 
\frac{1}{2} M^2\ \sum_n \int d^2x\, d^2y
\, \langle\bar\psi\psi(\omega)\bar\psi\psi(x)
\bar\psi\psi(y)\rangle_{M=0}^{(n)} + \dots
\label{cm2}
\eea

In terms of the chiral decomposition mentioned above, we write
\beq
\langle \bar\psi\psi(x)\rangle =\langle \bar\psi_+\psi_+(x)
\rangle+ \langle\bar\psi_-\psi_-(x) \rangle
=\langle s_+(x)\rangle +\langle s_-(x) \rangle
\eeq
As we have seen, the existence of $n N$ zero modes in topological 
sector $n$ implies
the vanishing of the first summatory in eq.(\ref{cm2}) $\forall n$. 
However, for higher powers of $M$ it is clear that certain nonzero
contributions come into play. 
As we have explained, the existence of zero modes with
a definite chirality (positive for $n>0$, negative for $n<0$) set
the (Grassman) integration rules of v.e.v.'s.
Since the fermionic sector has been completely decoupled, the counting 
of the nonzero pieces simply follows from that of the Abelian case, 
because the topological structure here can be red out from the 
torus of the gauge group.
To be more specific: the first $N_c$ powers of $M$
($j= 1\dots N_c$) receive an input from (just) the trivial 
topological sector.
Then, for $j\geq N_c$, the $n=1$ sector starts contributing 
together with $n=0$.
For powers $j\geq 2N_c$  the contribution of topological 
sector $n=2$ starts, and the counting follows so on in this way.

Now, it can be easily seen that the number of contributions grows 
together with the number of colors.
Usually, one investigates the full quark condensate
$\langle \bar q q(x)\rangle_{M}$ with the color index summed over; 
then, one  has a sum of $N_c$ identical nonzero components 
$\langle \bar q^c q^c(x)\rangle_{M}$ so that the complete condensate
reads  $N_c \langle \bar q^c q^c(x)\rangle_{M}$ (here $c$ 
represents any color).
As we let $N_c$ go to infinity the elementary massive condensate does so.
On the other hand, since within each nontrivial topological sector
the number of zero modes grows also to infinity, one has
divergent v.e.v's everywhere in the series expansion of eq.(\ref{cm2}).
Accordingly, high order terms could also produce a nonzero outcome
in the chiral limit.

Therefore, it is clear that the limit $M\ra 0$ becomes matter of a
careful analysis; namely, combined with a large number
of colors,  eq.(\ref{cm2}) leaves place 
enough for a nontrivial elementary condensate even in the chiral limit.


\section{Summary and discussion}

V.e.v.s of an arbitrary number of
fermionic bilinears in multiflavor non-Abelian gauge 
theories in Euclidean two space-time dimensions have been presented
in order to discuss  chiral symmetry issues of massless {\em QCD$_2$}. 

By means of a path-integral procedure we have shown how topological effects
give rise to non trivial correlators for a theory with a vanishing
homotopy group,  $\pi_1 (SU(N_c))=0$. 
These results make apparent that the topological structure found 
in QCD$_2$ for matter in the fundamental representation 
is indeed very important. As we have
explained, gauge fields lying in the Cartan 
subalgebra of $SU(N_c)$ have to be taken into
account to find a significant outcome for fermion condensates
although in the massless theory the elementary condensate is identically
zero.

Let us mention
that, in contrast, the elementary mass like condensate is not zero
when massless fermions are in the adjoint representation of the 
gauge group; however it is not an order parameter for the chiral 
phase transition  and it does not breakdown any
continuos symmetry. The existence of this condensate is apparent 
from bosonization rules and it is also expected from its topological properties
as has been argued in section 2.
With a standard normalization this v.e.v. of Majorana
fermions ammounts to $\langle\bar\Psi^a\Psi^a\rangle \sim 
\sqrt{g^2 N_c} N_c$ for a vanishing fermion mass 
and $N_c$ not necessarily divergent \cite{smilga}. This value
coincides with the one obtained in the 't Hooft regime for fundamental
matter but in this case only for an infinite number of colors 
\cite{zhit,lenz}.

Notice that as $N_c\ra \infty$ the relevant homotopy group of $SU(N_c)$
aproaches $Z$. One could be tempted to conclude that this fact relates
the non-Abelian theory in the large $N_c$ limit to its Abelian counterpart, 
in the sense that the
anomalous chiral symmetry breaksdown in the singlet channel; however, 
the inclusion of a mass term  seems
to be crucial for just the non-Abelian case.

Our analysis of the massive theory, put forward 
a possible way out to the BKT phenomenon by means of a path-integral
approach and topological considerations. In this way, both the trivial result
in the massless theory and the nonzero outcome in the massive
one, have gained contact with those emerging from alternative 
procedures \cite{zhit,lenz,smilga}.
The discussion
emphasizes the important role of the path taken to the chiral phase limit
in order to obtain finite results.

In contrast to the alternative approaches mentioned above, our treatment 
makes apparent the crucial role of topology into the properties of
the QCD vacuum, which for this reason, is a suitable scheme 
to analyze this issue. 
In spite of the previous discussion, for any finite value of $N_c$ 
the axial anomaly in the multiplet channel
dinamically breaks the flavor symmetry of QCD$_2$
for fundamental matter, allowing the existence of 
nonvanishing correlators for a larger number of points, 
see e.g. eq.(\ref{nabex}). 
In particular, it implies the circumvention of a  
decomposition ansatz and the large $N_c$ approximation
becomes unnecessary in order to find nontrivial multipoint correlators.
The elementary (massless) condensate appears to be nontrivial just for an infinite
number of colors in the chiral limit of the massive theory, a result
which comes forth in our treatment too.
As we have shown, condensates are completely determined by the topological
structure of the theory; however,
the question of exactly which topologically nontrivial configuration 
or master field could saturate
the elementary condensate for $N_c =\infty$ remains unanswered in an
analytical way.

\section*{Acknowledgments}  The author is grateful to 
Centro Brasileiro de Pesquisas F\'\i sicas (CBPF) and CLAF-CNPq,
Brazil, for warm hospitality and financial support. 
J.~Stephany is acknowledged for enlightening discussions.
F.A.~Schaposnik is acknowledged for his constant advice.


\end{document}